\documentclass[amsmath,amssymb,aps,pre,twocolumn
]{revtex4-2}

\usepackage{graphicx}
\usepackage{float}
\usepackage{subcaption}
\usepackage{physics}
\usepackage{hyperref}
\usepackage{color}
\usepackage{xr}

\definecolor{Black}{rgb}{0.00, 0.00, 0.00}
\definecolor{Blue}{rgb}{0.00, 0.00, 0.80}
\definecolor{Red}{rgb}{0.80, 0.00, 0.00}
\definecolor{Green}{rgb}{0.00, 0.50, 0.00}
\definecolor{Purp}{rgb}{0.50, 0.00, 0.50}

\hypersetup{
    colorlinks=true,       
    linkcolor=red,          
    citecolor=blue,        
    filecolor=magenta,      
    urlcolor=cyan           
}

\begin{document}

\title{Minimizing the Profligacy of Searches with Reset}

\author{John C. Sunil}
\author{Richard A. Blythe}
\author{Martin R. Evans}
\affiliation{SUPA, School of Physics and Astronomy, University of Edinburgh, Peter Guthrie Tait Road, Edinburgh EH9 3FD, UK}

\author{Satya N. Majumdar}
\affiliation{Universit{\'e} Paris-Saclay, CNRS, LPTMS, 91405, Orsay, France}

\date{\today}

\begin{abstract}
We introduce the profligacy of a search process as a competition between its expected cost and the probability of finding the target. The arbiter of the competition is a parameter $\lambda$ that represents how much a searcher invests into increasing the chance of success. Minimizing the profligacy with respect to the search strategy specifies the optimal search. We show that in the case of diffusion with stochastic resetting, the amount of resetting in the optimal strategy has a highly nontrivial dependence on model parameters resulting in classical continuous transitions, discontinuous transitions and tricritical points as well as non-standard discontinuous transitions exhibiting re-entrant behavior and overhangs.
\end{abstract}

\maketitle

\section{Introduction}
The task of searching arises in numerous domains. In nature, examples range from proteins locating their binding sites within the cell \cite{BN13} to foraging by macro-organisms \cite{Stanley,Bell}. In computer science, search algorithms have long been of fundamental interest \cite{Knuth} and have gained cultural importance in determining how the large and unstructured body of data that constitutes the internet is experienced by users \cite{Henzinger07}. Crossovers between these domains also exist, such as biologically-inspired optimization algorithms applied to solve many and varied science and engineering problems \cite{Darwish18}.  A large literature  has established optimal random search strategies, in the sense of optimizing the efficacy of the search \cite{VBHLRS99,BCMSV05,BC09,BLMV11,fronhofer2013random}.

In this work, we study a feature that is common to all such search processes, namely that increasing efficacy typically comes with a cost, for example, the amount of time or energy that must be invested or the complexity of the algorithm. A natural question is whether one can identify a  point 
of diminishing return, i.e., a point 
beyond which investing more effort into the search is not compensated by sufficiently increased success. We answer this question by introducing a quantity called \emph{profligacy} that expresses the cost-efficacy trade-off in a manner similar to Helmholtz free energy, {wherein the search cost plays the role of energy and the success probability provides an analog to entropy. The arbiter of the competition 
 is a temperature-like quantity $\lambda$, which has  units of cost, and characterizes how much extra one is prepared to invest in order to increase the success probability of a search.}
 
 We will show below that in the context of even fairly simple diffusive searches,  namely Resetting Brownian Motion \cite{EM11a,EM11b,EM14,EMS20} and Resetting Brownian Bridge \cite{BMS22}, the optimal strategy that arises from minimizing the profligacy exhibits a rich phase diagram (presented in Fig.~\ref{fig:transitions} below). By this we mean that one finds both continuous and discontinuous transitions between regimes in which the optimal search strategy switches
 from a strategy  with no resetting to one with a finite resetting rate.
 Moreover, the phase structure goes beyond what is normally seen at equilibrium, exhibiting re-entrant behavior and overhangs, to be detailed below.

  The paper is organized as follows. In Section~\ref{sec:prof}
 we  define and motivate the profligacy as a function that captures the cost-efficacy trade-off of a search process. In Section~\ref{sec:searches} we define the resetting search processes that we consider. In Section~\ref{sec:cs} we present expressions for the average cost and success probabilities of these searches. In Section~\ref{sec:min} we present phase diagrams obtained by minimizing the progligacy and in Section~\ref{sec:landau} we present a Landau-like theory that explains the features of these phase diagrams. Finally, we conclude in Section~\ref{sec:sum}.
\begin{figure}
  \centering
    \includegraphics[width = 0.4\textwidth]{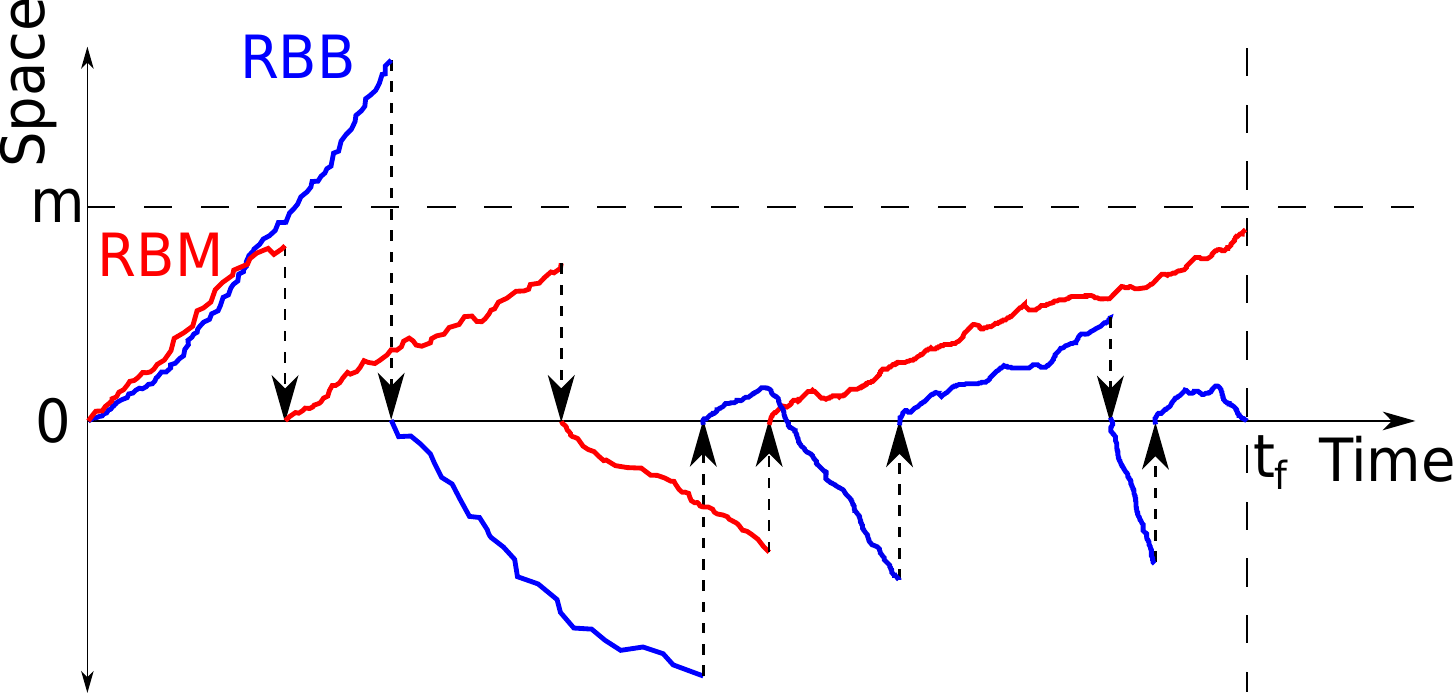}
    \caption{
    Schematic trajectory for RBM (red) and RBB (blue). For RBM, the particle   diffuses without constraint  while for RBB the particle is constrained to return to the starting position at the completion time $t_f$. The vertical arrows represent resetting events. The searcher is considered to have found the target if it crosses $x=m$ in any of its excursions before $t_f$.
    }
    \label{fig:schem}
\end{figure}
\section{Profligacy of the search process}\label{sec:prof}
 In this section we define the profligacy of the search.
To be concrete, let us define the set-up we have in mind. Imagine a search company with $K$ independent searchers each of whom is following a search strategy that is controlled by a search parameter $r$, for example
it could be Brownian motion with resetting to the origin at rate $r$, see Fig.~\ref{fig:schem}. In addition the search has an allotted time $t_f$.
Such  fixed duration search processes naturally appear in many contexts e.g. when a drone recording data is hired for a fixed time  or when a search  helicopter with a fixed amount of fuel has a limited flight time to return to its base.
Associated with the trajectory of each searcher $j$  is a non-negative cost $C_j$. 
The company also has a  budget $C_b$ per searcher.
For a given realisation of the full search process (trajectories of all $K$ searchers), this budget implies a hard constraint $\sum_{j=1}^K C_j/K = C_b$. We assume that $K$
is very large, such that
$\sum_{j=1}^K C_j/K \to \langle C(r)\rangle$, where angled brackets denote the average
of the cost over the trajectories of a `single' searcher. In this limit we have an effective single searcher problem.

A fixed budget $C_b$ immediately
constrains the value of the search parameter to be $r_1$ where $\langle C(r_1)\rangle= C_b$.
For simplicity, we assume there is a unique value of $r_1$, and that all the searchers start from the origin in a one-dimensional setting.
Now imagine that we have the task of finding a fixed target at distance $m$ from the origin, see Fig.~\ref{fig:schem}. The search has to complete the allotted time $t_f$ even if the target is located before. For example, an automated drone will continue recording data, to be analysed afterwards,  until $t_f$. Also, the limited search time $t_f$ means there is a finite probability that a searcher may not find the target. We define $P_s(r)$ as the success probability, that is, the probability that a searcher locates the target before $t_f$.

The goal is to maximize the success probability by choosing an appropriate value of $r$.
However, a fixed budget constraint only allows the value $r_1$ and $P_s(r)$ is typically not maximal at $r_1$, but at $r_2\neq r_1$.  This leads to the dilemma of how to maximize $P_s(r)$ and at the same time respect budgetary restrictions. This requires one to soften the hard budgetary constraint $\delta( \langle C(r)\rangle -C_b)$.
A natural way to implement this softening is to replace the delta function by   
an exponential (as is standard in statistical mechanics when one goes from the microcanonical to canonical ensemble by replacing a delta function by an exponential Boltzmann weight).  Hence, we consider  a weighted  efficacy $P_{\rm s}(r) e^{-(\langle C(r)\rangle-C_b)/\lambda}$ where  $\lambda$ is a temperature-like softening parameter in analogy with a Boltzmann weight with $\langle C(r)\rangle -C_b$ playing the role of
energy. Note that the choice of exponential as the softening function is quite natural.
In fact this is the standard procedure in statistical mechanics when one goes from microcanonical to canonical ensemble and  replaces a delta function constraint by an exponential weight. However, other choices for the softening function are  also possible e.g. a Gaussian centred at $C_b$.  We expect that the results and conclusions will not change qualitatively, although quantitative details will depend on the choice. In the limit $\lambda \to 0$ the  hard constraint is recovered and $r$ is restricted to the value $r_1$. In the opposite limit $\lambda \to \infty$, the constraint on budget disappears and the optimal parameter value  is $r_2$ at which $P_s(r)$ is maximized.
By varying $\lambda$ one interpolates between the two limits.

We call $\lambda$  the investment, as it has units of cost and characterises how much extra a client is prepared to pay to improve the chance of success.
The weighted efficacy  can then be expressed as $e^{-(\xi-C_b)/\lambda}$
where the {\em profligacy} $\xi$ is given by
\begin{equation}
    \xi = \langle C(r) \rangle - \lambda \ln P_{\rm s}(r) \;.
    \label{profligacy}
\end{equation}
Now the objective is to find the value of $r$ that minimizes the profligacy.
This is somewhat reminiscent of a Helmholtz free energy where $\langle C(r) \rangle$
plays the role of energy while $\ln P_{\rm s}(r)$ plays the role of entropy.
The profligacy  \eqref{profligacy} is a single quantity that allows one to optimize the trade-off between the competing objectives  of maximizing efficacy and minimizing cost.

\section{Resetting Search Strategies}
\label{sec:searches}
We devote the rest of this work to determining 
how the search strategy that minimizes the profligacy changes as we vary the investment $\lambda$, within the framework of diffusive searches under stochastic  resetting. In this context the searcher is modeled as a diffusive particle starting from the origin, which is reset instantaneously to the origin with rate $r$ \cite{EM11a}, see Fig.~\ref{fig:schem}. The search is successful if the searcher reaches a target located at distance $m$ from the origin. Early studies of such processes \cite{EM11a,EM11b,EMM13,EMS20} demonstrated how resetting allows the target to be found more quickly than through diffusion alone, replacing an infinite mean time to locate the target with some finite value. Moreover, there exists an optimal resetting rate $r^*$, which minimizes the mean time to find a target \cite{EM11a} and the value of $r^*$ undergoes phase transitions as various control parameters are varied \cite{KMSS14,CM15,CS15,Reuveni16,BBR16,PR17,CS18,RMR19,PP19b,MBMS20,SB21,MBM22,BMS23,BMMS23}.  
 Recently, the consequences of associating a cost with each reset, accounting for the consumption of time, fuel or some other finite resource, have been investigated \cite{PKR20,BS20a,GP22,DBM23,SBEM23,MOK23,OGMK23,MM23,GLPHMG23,PPPL23,OG24} and the statistics of the cost as a function of $r$ have been computed \cite{SBEM23}. 
{ We emphasize that the  problem we consider here is different, in that  we consider a predetermined time and cost of the search.
For example, a client purchases a search period $t_f$ regardless of whether the target is found within that time. 
 We adopt the fixed-time  time search constraint  as it provides a broad framework in which different ensembles of trajectories and different cost functions can be considered.
In all these cases the resetting rate $r$ furnishes a single search parameter
with which we may optimize the profligacy.} We will focus on  transitions from a non-resetting optimal strategy, $r=0$, to a non-zero optimal value of $r$.

We consider two classes of resetting models which differ in what happens at the end time $t_f$, see Fig.~\ref{fig:schem}. In the case of Resetting Brownian Motion (RBM) \cite{EM11a}, the process is simply halted at $t_f$,  meaning that searcher
can be anywhere in space at time $t_f$.
By contrast, a Resetting Brownian Bridge (RBB) \cite{BMS22} imposes the additional constraint that a searcher must return to the origin at time $t_f$. This models such situations as a rescue helicopter having to return to base to refuel after a prespecified flight time. The RBB ensemble is obtained from that of RBM by retaining only those trajectories that occupy the starting position at the completion time, $t_f$. In mechanical terms, this conditioning creates a time-dependent drift on the particle motion along with a resetting rate that diverges as $t_f$ is approached, thereby  guaranteeing a return to the origin \cite{BMS22}.

\section{Cost and Success Probability Calculations}\label{sec:cs}
Let us define the class of cost functions that we consider for the  resetting searches.
Each reset $i$ contributes a cost $c_i \ge 0$  which depends on the distance travelled to the origin in the reset. 
 The number of resets $N$ that occur up to the
fixed time $t_f$ is
a random variable and fluctuates from trajectory to trajectory.
The total cost, $C$, of a trajectory is obtained by summing over all $N$ resets that occur along it:
\begin{equation}
    C= \sum_{i=1}^{N}c_i= \sum_{i=1}^{N} c(|y_i|)\label{C}
\end{equation}
 and the cost is zero if there is no reset in the trajectory.
Here, $y_i =x_i/\sqrt{2D t_f}$ is the rescaled (dimensionless) position just before the reset and the function $c(y)$ is the cost per reset.  Similarly,  it is convenient to use dimensionless, rescaled variables $R =rt_f$ and $M=m/\sqrt{2Dt_f}$, which  eliminate the values of the diffusion constant $D$ and  search time $t_f$ from the discussion, and leave  $R$, $\lambda$ and $M$ as the control variables. We will compare a linear, $c(y) = \sqrt{2} y$, and a quadratic, $c(y) = 2y^2$, cost per reset, since these are the simplest cases for practical applications, but, as we shall see, yield very different phase diagrams. The linear cost can be motivated as the time required to bring the particle back to the origin at a constant velocity \cite{PKR20,BS20a,BS20b}. Likewise, the quadratic cost can represent energy consumption, monetary cost or thermodynamic cost for the particle to reset \cite{FPSRR20,BBPMC20,MOK23}. { We note that other kinds of costs could be relevant to a search process, for example those dependent on the time of resets, distance traveled between excursions, or the number of previous resets {\it etc.} However, for simplicity we leave other choices for future studies.}

Our aim now is to determine the optimal resetting strategy---that is, the value of 
$ R= R^*$ that minimizes the profligacy (\ref{profligacy})---for a given combination of target position $M$ and investment $\lambda$. To achieve this, we must first evaluate the mean cost $\expval{C}$ and the probability of finding the target $P_{\rm s}$. The resetting systems we consider have the appealing feature that these quantities can be calculated analytically, following recent progress in leveraging renewal properties of the process \cite{SBEM23}.   The success probability has been calculated for RBM in \cite{EM11a} and for RBB in \cite{BMS22}. The mean costs are derived in detail in the Appendix~\ref{sec:cost} and the details of the computation of success probabilities is discussed in Appendix~\ref{sec:sprob}.

\begin{figure*}
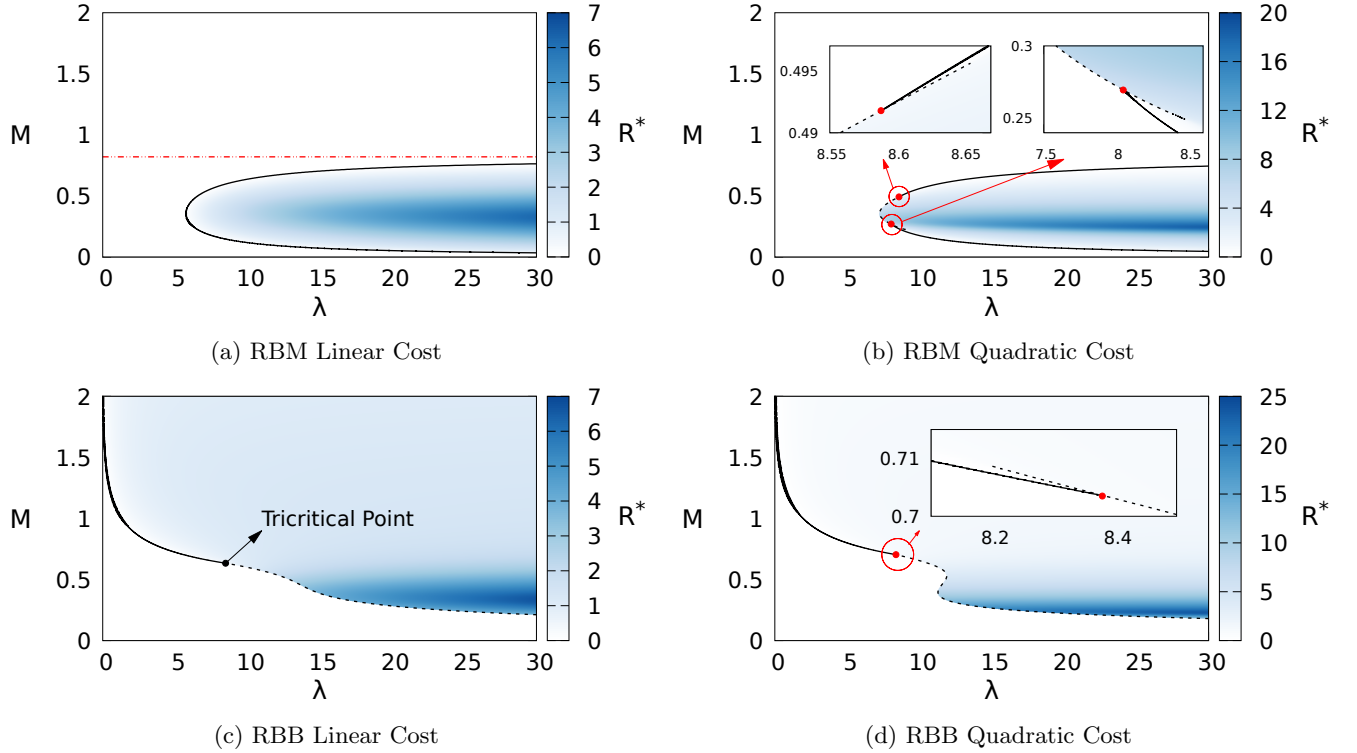

\centering
    \begin{subfigure}{.49\textwidth}
		\centering
	    \includegraphics[width=\linewidth]{RBM_linear_profligacy_transitions.pdf} 
		\caption{RBM Linear Cost}
        \label{fig:RBM_linear}
	\end{subfigure}
    \begin{subfigure}{.49\textwidth}
		\centering
	    \includegraphics[width=\linewidth]{RBM_quadratic_profligacy_transitions.pdf} 
		\caption{RBM Quadratic Cost}
        \label{fig:RBM_quad}
	\end{subfigure}
	\begin{subfigure}{.49\textwidth}
		\centering
	       \includegraphics[width=\linewidth]{RBB_linear_profligacy_transitions.pdf} 
        \caption{RBB Linear Cost}
        \label{fig:RBB_linear}
		\end{subfigure}
		\begin{subfigure}{.49\textwidth}
		\centering
	    \includegraphics[width=\linewidth]{RBB_quadratic_profligacy_transitions.pdf} 
		\caption{RBB Quadratic Cost}
        \label{fig:RBB_quad}
	\end{subfigure}
 \caption{Phase diagrams: heat map of $R^*$ which minimizes profligacy $\xi$ \eqref{profligacy} in the $M$--$\lambda$ plane for 4 different cases: (a) RBM with linear cost (b) RBM with quadratic cost (c) RBB with linear cost and (d) RBB with quadratic cost. The phase boundaries delineate the regions of zero and non-zero $R^*$:  a full line indicates a continuous  transition and broken line a discontinuous transition. In (a) the  horizontal red line { (obtained from the Landau-like expansion \eqref{landau_expansion})} indicates the threshold, $M_{\rm T}$, above which there is no transition. { The red dots in (b) and (c) (also obtained from \eqref{landau_expansion}) indicate a transition of a different nature from the classical tricritical point, as described in the text.}  {The figures are obtained by evaluating $\xi$ for an equally spaced set of values of $R$ on a grid of $M$ and $\lambda$. $R^*$ is the value 
 of $R$ that minimizes $\xi$. If the value of $R^*$ goes from $0$ to the minimum non-zero $R$ value (chosen as 0.01) between $M$--$\lambda$ grid points, the transition  is deemed continuous; if the change is 5 times the minimum $R$ value, it is deemed discontinuous. }}
 \label{fig:transitions}
\end{figure*}
For the case of RBM, the explicit expressions are
 \begin{align}
     \expval{C}^{\text{RBM}}_{\text{lin}} = &
     \frac{{\rm e}^{-R}}{\sqrt{\pi }}+\frac{(2R-1)\erf\left(\sqrt{R}\right)}{2 \sqrt{R}} \label{lin_RBM}\\
     \expval{C}^{\text{RBM}}_{\text{quad}} = &
     \frac{2 \left(R+{\rm e}^{-R}-1\right)}{R} \label{quad_RBM}\\
      P^{\text{RBM}}_{\rm s} = &\int_{\Gamma} \frac{{\rm d}u}{2\pi i} {\rm e}^{u} \frac{1}{u} \frac{R+u}{R+u {\rm e}^{M\sqrt{2(u+R)}}} \label{phit_RBM}
 \end{align}
where the subscripts ${\rm lin}$ and ${\rm quad}$ refer to the linear and quadratic cost functions, respectively. The success probability is expressed as an inverse Laplace transform, denoted by an integral over the Bromwich contour $\Gamma$. This form is sufficient to determine the phase diagrams numerically using a suitable inversion algorithm \cite{AW06}.
For RBB, meanwhile, we obtain 
\begin{align}
     \expval{C}^{\text{RBB}}_{\text{lin}} &= 
     R\phi(R) \label{lin_RBB}\\
     \expval{C}^{\text{RBB}}_{\text{quad}} &= 
     2 - \frac{\erf(\sqrt{R})}{\sqrt{R}} \phi(R)\label{quad_RBB}\\
     P^{\text{RBB}}_{\rm s} &= \phi(R) \int_{\Gamma} \frac{{\rm d}u}{2\pi i} \frac{{\rm e}^{u} \sqrt{u+R}}{u}  \frac{R+u {\rm e}^{-M\sqrt{2}\sqrt{u+R}}}{R+u {\rm e}^{M\sqrt{2}\sqrt{u+R}}} \label{phit_RBB}
\end{align}
where $\phi(R) = \sqrt{\pi}\left[e^{-R}+\sqrt{\pi R}\erf\sqrt{R}\right]^{-1}$.

\section{Minimizing Profligacy}\label{sec:min}
 Having analytical expressions for the average cost and the success probability, we can now write down the profligacy using \eqref{profligacy} and the task at hand is to minimize the profligacy. In Fig.~\ref{fig:transitions} we present the phase diagrams in the $\lambda$--$M$ plane obtained by minimizing the profligacy with respect to $R$ for  the two different types of search (RBM and RBB) and the two different cost functions (linear and quadratic). In the unshaded regions, diffusing without resetting  is optimal ($R^*=0$), whereas in the shaded regions, a nonzero resetting rate yields the least profligate search.  Along the solid lines, the optimal resetting rate $R^*$ changes continuously across the phase boundary, whilst along the broken lines the optimal resetting rate jumps discontinuously. 
 The nature of the transition is significant: a continuous transition implies
that a gradual introduction of resetting yields the optimal search strategy whereas a discontinuous transition implies a sudden switch of strategies to a finite resetting rate.

As we now discuss, the significant differences in the topology of the four phase diagrams derive from small qualitative distinctions in the behavior of $\langle C \rangle$ and $P_{\rm s}$, i.e.~Eqs.~(\ref{lin_RBM})--(\ref{phit_RBB}).
\begin{figure*}
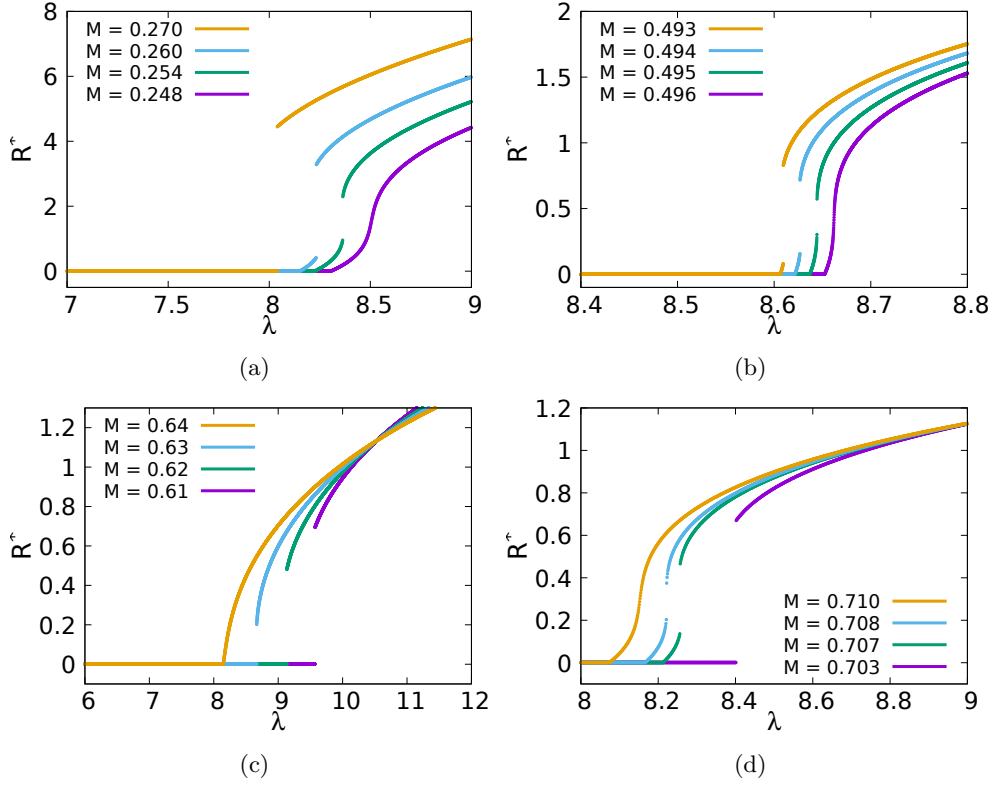

\centering
    \begin{subfigure}{.36\textwidth}
		\centering
	    \includegraphics[width=\linewidth]{RBM_tricritical_transition_1.pdf} 
		\caption{}
        \label{fig:RBM_overhang_transition_1}
	\end{subfigure}
    \begin{subfigure}{.36\textwidth}
		\centering
	    \includegraphics[width=\linewidth]{RBM_tricritical_transition_2.pdf} 
		\caption{}
        \label{fig:RBM_overhang_transition_2}
	\end{subfigure}
    \begin{subfigure}{.36\textwidth}
		\centering
	    \includegraphics[width=\linewidth]{RBB_tricritical_transition.pdf} 
		\caption{}
        \label{fig:tricritical_transition}
	\end{subfigure}
    \begin{subfigure}{.36\textwidth}
		\centering
	    \includegraphics[width=\linewidth]{RBB_overhang_transition.pdf} 
		\caption{}
        \label{fig:overhang_transition}
	\end{subfigure}
\caption{Optimal resetting rate $R^*$ versus $\lambda$ for values of $M$ in the regions where continuous and discontinuous transitions are in close proximity. Panels (a) and (b) RBM with quadratic cost: there are two different ranges of $M$ (see Fig.~\ref{fig:transitions}) where on increasing $\lambda$, there is first a continuous transition followed by a discontinuous transition. The discontinuous jump in $R^*$ closes at a non-zero value of $R^*$ as $M$ is varied and thus is not a usual tricritical point. Panel (c) RBB with linear cost: we have a usual tricritical point where the jump in discontinuity closes at $R^* = 0$. Panel (d) RBB with quadratic cost: similar to RBM the jump closes at non-zero value of $R^*$.}
 \label{fig:tricritical_overhang}
\end{figure*}
 The easiest phase diagram to understand is that for RBM and a linear cost per reset (Fig.~\ref{fig:RBM_linear}) for which the cost (\ref{lin_RBM}) increases monotonically with the resetting rate (see SM Fig.~\ref{fig:costs}).
When the target is far from the origin (large $M$), the success probability (\ref{phit_RBM}) monotonically decreases (see SM Fig.\ref{fig:ps}).
Thus for large $M$, $R^*=0$ is always the optimal value. However for $M$ below a threshold value $M_T$,  the success probability initially increases with $R$ and has a peak at some intermediate value of $R$.
This implies that for sufficiently large $\lambda$ the optimal resetting rate $R^*$ is non zero. The transition to an optimal strategy involving resetting
can be understood by appealing to a Landau-like theory, to be discussed in Section \ref{sec:landau}, which implies a classical continuous phase transition into a resetting phase ($R^*>0$) as $\lambda$ is increased at fixed $M<M_T$.

 We now turn to the case of a quadratic cost per reset,  for which the mean cost no longer grows without bound as $R\to\infty$ but approaches a plateau (see Appendix~\ref{sec:cost} Fig.~\ref{fig:costs}). The effect of this on the RBM phase diagram, Fig.~\ref{fig:RBM_quad}, is the addition of a discontinuous transition line at intermediate $M$. Although this meets the continuous transition line at two points, it does not end there (as at tricritical points) but \emph{extends} beyond them, thus creating overhangs (see insets). So as $\lambda$ is increased, for particular choices of $M$, we find an initial continuous transition to a nonzero optimal resetting rate $R^*$ very closely followed by a discontinuous jump in $R^*$ as shown in Figs.~\ref{fig:RBM_overhang_transition_1} and \ref{fig:RBM_overhang_transition_2}. This sequence of transitions implies an initial gradual introduction of resetting into the optimal search strategy then a sudden jump to a stronger resetting strategy.

 In contrast to RBM, the success probability for a RBB search is a peaked function of $R$ for all target positions $M$ (see SM Fig.\ref{fig:ps}). The effect of this is that resetting always becomes beneficial for high enough $\lambda$.
 With a linear cost per reset (Fig.~\ref{fig:RBB_linear}), the transition is continuous for high $M$ and discontinuous at low $M$, the two lines meeting at a classical tricritical point (see Fig.~\ref{fig:tricritical_transition}) \cite{BMMS23}. Finally, for the case of RBB with quadratic cost per reset (Fig.~\ref{fig:RBB_quad}), the shape of the phase boundary has developed a kink in comparison to the  linear cost case. Similar to the case of RBM, the mean cost plateaus as $R\to\infty$, creating an overhang instead of a tricritical point. This overhang effect is evident in Fig.~\ref{fig:overhang_transition}, where we see both a continuous transition from $R^*=0$ to $R^*>0$ and a discontinuous jump between two nonzero values of $R^*$ at a higher value of $\lambda$. As $M$ is increased, the jump in the discontinuous transition goes to zero and we are left with a single continuous transition. If we  fix $\lambda$ and increase $M$, then near to the kink we have re-entrant behaviour into the $R^*=0$ phase via a discontinuous transition.

\section{Landau-like Expansion}
\label{sec:landau}
To gain deeper insight into the nature of these transitions, we make  a Landau-like expansion of the profligacy \eqref{profligacy} in powers of $R$, valid for small values of $R$, 
 \begin{align}
   \xi = a_0 + a_1 R + a_2 \frac{R^2}{2} + a_3 \frac{R^3}{6} + a_4 \frac{R^4}{24} + \ldots\;, \label{landau_expansion}
   \end{align}
with all the coefficients $a_i = a_i(\lambda,M)$. Due to the absence of
an $R \to -R$ symmetry, we have to include all terms of the expansion. {Related Landau-like expansions have been made in \cite{PP19b,BMMS23,JSDB24}. The expansion of $\langle C\rangle$ is obtained directly from \eqref{lin_RBM}, \eqref{quad_RBM} and \eqref{lin_RBB}, \eqref{quad_RBB}. We also require the expansion of $P_{\rm s}$ for  the two resetting ensembles, obtained by expanding out the integrands of \eqref{phit_RBM} and \eqref{phit_RBB} in terms of $R$ and integrating term by term. The details of the expansion for $P_S$ are provided in Appendix~\ref{appendix_landau}.

We now discuss the different cases that occur for transitions in the global minimum of the profligacy $\xi$ as we vary $\lambda$.

\subsection{Classical Continuous Transition}
The simplest case is when $a_2 > 0$ in the expansion \eqref{landau_expansion} and  we may ignore higher order terms.  The curve of continuous transitions, $\lambda^*(M)$, is obtained by solving $a_1(\lambda,M) = 0$ for $\lambda$,  as for a continuous transition seen in equilibrium systems \cite{Goldenfeld}. This simple scenario pertains in the case of RBM with a linear cost, where we saw in Fig. \ref{fig:transitions} that a continuous transition occurs on increasing $\lambda$, for sufficiently low $M$. 
\begin{figure}[tb]
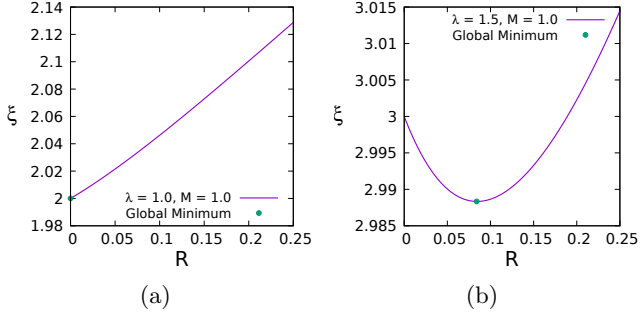

\centering
    \begin{subfigure}{.235\textwidth}
		\centering
	    \includegraphics[width=\linewidth]{RBB_Landau_classical_continuous_1.pdf} 
		\caption{}
        \label{fig:c_c_1}
	\end{subfigure}
    \begin{subfigure}{.235\textwidth}
		\centering
	    \includegraphics[width=\linewidth]{RBB_Landau_classical_continuous_2.pdf} 
		\caption{}
        \label{fig:c_c_2}
	\end{subfigure}
 \caption{The figure presents classical continuous transition observed in $\xi$ for the case of RBB with linear cost per reset when $\lambda$ is varied for $M=1.0$. (a) The global minimum which is at $R^* = 0$ initially continuously transitions to the new global minimum $R^*$ in (b) as $\lambda$ is varied.}
 \label{fig:c_c}
\end{figure}

For the case of RBM the continuous transition condition,  $a_1(\lambda,M) = 0$, is satisfied at $\lambda^*(M)$  given by
\begin{align}
    \lambda^* =  \frac{d\expval{C}^\text{RBM}}{dR}\bigg\vert_{R \to 0^+}\times\frac{\text{erfc}\left( \frac{M}{\sqrt{2}} \right)}{g_1(M)}\;, \label{RBM_continious_transition}
\end{align}
with
\begin{align}
    \frac{d\expval{C}^{\text{RBM}}_{\text{lin}}}{dR}\bigg\vert_{R \to 0^+} = \frac{4}{3\sqrt{\pi}}\;,
\end{align}
and
\begin{align}
    \frac{d\expval{C}^{\text{RBM}}_{\text{quad}}}{dR}\bigg\vert_{R \to 0^+} = 1\;.
\end{align}
However, since we consider the investment $\lambda$, to act as a penalty if the searcher does not find the target, we require $\lambda>0$. But $g_1(M)<0$ for $M>0.8198\dots$ which suggests that a continuous transition cannot exist for RBM if the rescaled distance to the target is greater than the threshold value $M_T = 0.8198\dots$. Further, $a_1(\lambda^*,M)>0$ for all values $M<M_T$, which implies that the system has a continuous transition for all values of $M<M_T$. {At the threshold value, $\lambda^*(M_{\rm T})\to \infty$ and for $M>M_{T}$ no transition occurs. The exact value $M_{\rm T}= 0.8198\ldots$ that is obtained from this procedure provides the horizontal line in Fig.~\ref{fig:RBM_linear}.} This exactly matches with the result obtained in the FIG.~\ref{fig:RBM_linear} and \ref{fig:RBM_quad} where no transitions are observed beyond the threshold value, $M_T$. 

For the case of RBB, the continuous transition condition,  $a_1(\lambda,M) = 0$, is satisfied at $\lambda^*(M)$. There we obtain the curve of continuous transition ($a_1(\lambda, M)=0$) to be
\begin{align}
    \lambda^* = \frac{d\expval{C}}{dR}\bigg\vert_{R \to 0^+}\times \frac{e^{-2M^2}}{h_1(M)}\;, \label{RBB_continuous_transition}
\end{align}
where we have the  derivatives
\begin{align}
    \frac{d\expval{C}^{\text{RBB}}_{\text{lin}}}{dR}\bigg\vert_{R \to 0^+} = \sqrt{\pi} \;,
\end{align}
and
\begin{align}
    \frac{d\expval{C}^{\text{RBB}}_{\text{quad}}}{dR}\bigg\vert_{R \to 0^+} = \frac{8}{3}\;.
\end{align}
 Since $h_1(M)>0$ for all values of $M$, we obtain a solution for $a_1 =0$ for all values of $M$ for RBB.
For small values of $\lambda^*$ this corresponds to a continuous transition, however for large values of $\lambda^*$ we find  that $a_2<0$ and we have to consider a classical discontinuous transition which we now discuss.

\subsection{Classical Discontinuous Transition}
The next case is where $a_2$ may be positive or negative according to parameters, but $a_3$ is positive. This is the case for RBB  with linear cost, Fig.~\ref{fig:RBB_linear}. For $a_2>0$ we obtain a continuous transition at $a_1=0$ but for $a_2<0$ there is a discontinuous transition.
We then have a local minimum in the profligacy at a non-zero value of $R$ in addition to a boundary minimum at $R=0$. If on increasing $\lambda$ the  global minimum switches between these two local minima, we have a jump in $R^*$ from a zero to a non-zero value. This is the classical scenario for a discontinuous transition. A classical tricritical point  occurs when a continuous transition line meets a discontinuous transition line, which occurs when $a_1=a_2=0$.
\begin{figure}[tb]
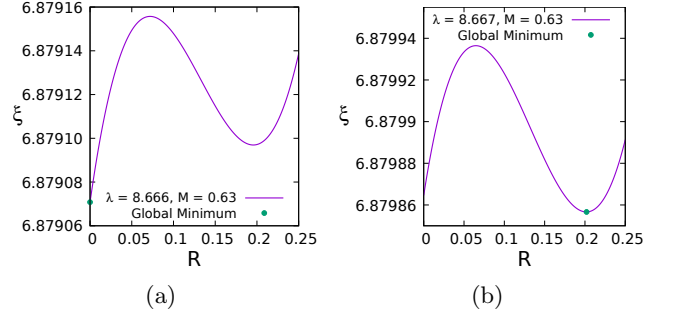

\centering
    \begin{subfigure}{.235\textwidth}
		\centering
	    \includegraphics[width=\linewidth]{RBB_Landau_classical_discontinuous_1.pdf} 
		\caption{}
        \label{fig:c_dc_1}
	\end{subfigure}
    \begin{subfigure}{.235\textwidth}
		\centering
	    \includegraphics[width=\linewidth]{RBB_Landau_classical_discontinuous_2.pdf} 
		\caption{}
        \label{fig:c_dc_2}
	\end{subfigure}
 \caption{The figure presents classical discontinuous transition observed in $\xi$ for the case of RBB with linear cost per reset when $\lambda$ is varied for $M=0.63$. (a) The global minimum which is at $R^* = 0$ initially discontinuously transitions to the new global minimum $R^*$ in (b) when the local minima becomes the new global minima as $\lambda$ is varied.}
 \label{fig:c_dc}
\end{figure}

A classical tricritical point occurs in the case of RBB  with linear cost as is seen in Fig. \ref{fig:RBB_linear}. Then for $a_2>0$ we obtain a continuous transition at $a_1=0$ but for
$a_2<0$ there is a discontinuous transition and if $a_1 = a_2 = 0$ (and $a_3>0$), we obtain a  tricritical point.  To obtain the tricritical point for RBB with  linear cost per reset, we set $a_2(\lambda^*,M) = 0$, which gives
\begin{align}
     e^{-2M^{*2}}\frac{d^2 \expval{C}}{dR^2}\bigg\vert_{R \to 0^+} + e^{2M^{*2}}\lambda^* h_1^2(M^*)
    -\lambda^* h_2(M^*) = 0\;,
\end{align}
which upon solving yields the tricritical point $(M^*,\lambda^*) = (0.6333,8.4994)$ (filled circle in Fig.~\ref{fig:RBB_linear}). For $\lambda$ approaching this point from above we have a classical discontinuous transition

\subsection{Non-classical Discontinuous Transition}

 The interesting non-classical overhangs that occur for both RBM and RBB with quadratic cost per reset (Figs.~\ref{fig:RBM_quad} and \ref{fig:RBB_quad}) can be attributed to the coefficients satisfying $a_3<0$ and $a_4>0$. If we look for a tricritical point by solving $a_2(\lambda^*,M) = 0$ for $M$, we find at all such $M^*$ that $a_3(\lambda^*,M^*)<0$. This violates the condition for a tricritical point, which is why overhangs emerge instead. 

This non-standard scenario occurs when $a_3<0$ and $a_4>0$ and we retain terms up to $R^4$ in the Landau-like expansion. This is the case for both  RBM and RBB with quadratic cost per reset in an intermediate range of $M$. The quartic  expansion in $R$ allows two local minima in the profligacy if $a_1<0$,
or one local minimum, plus a boundary minimum at $R=0$, if $a_1>0$.  In the latter case either a discontinuous transition to a non-zero value of $R^*$ can occur when the local minimum becomes the global minimum, or a continuous transition can occur when $a_1$ becomes negative; in the former case a discontinuous transition between two non-zero values of $R^*$ can occur when the global minimum switches between the two local minima. A sequence of a continuous transition (when $a_1$ turns negative) followed by a discontinuous transition (when the global minimum switches between the subsequent two local minima) generates the overhangs seen in Figs. \ref{fig:RBM_quad},\ref{fig:RBB_quad}. Thus the red dots in Figs. \ref{fig:RBM_quad},\ref{fig:RBB_quad}
are not classical tricritical points as can be verified by evaluating $a_2(\lambda^*,M) = 0$.
For RBM evaluating this results in two different values $M^*_{1,2}$, but evaluating the coefficient of $R^3$ gives $a_3(\lambda^*,M^*_{1,2})<0$. 

\begin{figure}[tb]
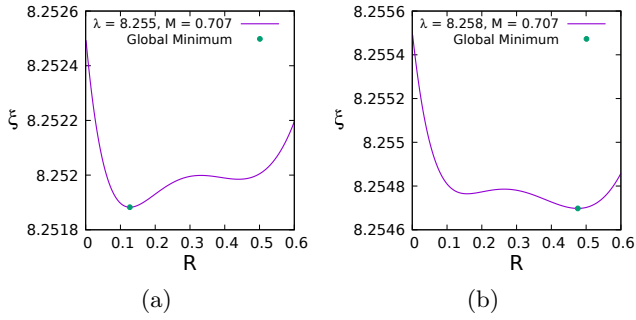

\centering
    \begin{subfigure}{.235\textwidth}
		\centering
	    \includegraphics[width=\linewidth]{RBB_Landau_non_classical_discontinuous_1.pdf} 
		\caption{}
        \label{fig:nc_dc_1}
	\end{subfigure}
    \begin{subfigure}{.235\textwidth}
		\centering
	    \includegraphics[width=\linewidth]{RBB_Landau_non_classical_discontinuous_2.pdf} 
		\caption{}
        \label{fig:nc_dc_2}
	\end{subfigure}
 \caption{The figure presents the non-classical discontinuous transition occurs observed in RBB with quadratic cost per reset as $\lambda$ is varied for $M = 0.707$. (a) The global minimum initially emerges continuously from $R = 0$ as $\lambda$ is varied. (b) As lambda is further increased, the second minimum becomes the global minima and there is a discontinuous transition of the optimal value $R^*$ to the new global minimum.}
 \label{fig:nc_dc}
\end{figure}
{The behaviours that we have obtained 
within the Landau-like expansion and with linear and quadratic cost functions are generic. That is, we have explored all the possible behaviours coming from the different values $a_i$'s could take, for $i=1\ldots 3$ with $a_4$ positive.
To obtain different phenomenology  would require cost functions  where higher coefficients have to be taken into account, and possibly more exotic transitions could be obtained.} Of course the predictions of a Landau-like theory for discontinuous transitions will only  be quantitatively accurate for small $R$, nevertheless
the correct qualitative behaviour is predicted.

\section{Summary and Outlook}
\label{sec:sum}

In summary, we have introduced the profligacy $\xi$ \eqref{profligacy} as a tool to analyse the cost-efficacy 
trade-off in a search process. We have derived $\xi$ from considering the efficacy, $P_s$, weighted by an exponential function of the cost expectation value $\langle C \rangle$. The simple framework of diffusion under stochastic resetting with rate $r$ has allowed us to derive analytical expressions for $P_s$ and $\langle C \rangle$ and thus to carry out the minimization of $\xi$. This has resulted in surprisingly rich phase diagrams,
exhibiting classical continuous and discontinuous transitions,  but also non-standard transitions with re-entrant behaviour and overhangs. 
 These transitions imply changes of the optimal search strategy that may be gradual or sudden.  We have shown that these transitions may be understood within a simple Landau-like expansion of the profligacy.  

As the profligacy just requires a cost and a measure of success as inputs, it has the potential for application in wider contexts.
 So far we have considered resetting Brownian motion and resetting Brownian bridge searches where we are able to compute the profligacy analytically. The Landau-like theory implies that the observed transitions should be
generic. 
It would be of interest to see if the different classes of transition
that we have identified here, arise
more generally for other searches
and cost functions.

\begin{acknowledgments}
\textit{Acknowledgments}---For the purpose of open access, the authors have applied a Creative Commons Attribution (CC BY) licence to any Author Accepted Manuscript version arising from this submission.  JCS thanks the University of Edinburgh for  the award of an EDC Scholarship.
\end{acknowledgments}

\appendix

\section{Derivation of the Mean Cost}
\label{sec:cost}

The key quantity that is required to determine the mean total cost incurred in a diffusive resetting process is $\Omega_r(C,t_f)$, the statistical weight of trajectories  that start at the origin at $t=0$, return to the origin as a Poisson process at rate $r$, end at a predetermined time $t=t_f$ and incur a total cost $C$. These statistical weights will differ in the ensemble where the endpoint of the trajectory is free (as in Resetting Brownian Motion, RBM) or constrained to lie at the origin (as in a Resetting Brownian Bridge, RBB). Once this quantity is known, we can determine the mean cost over such trajectories as
\begin{equation}
    \expval{C} = \frac{\int{\rm d}C\, C\, \Omega_r(C,t_f)}{\int{\rm d}C\,\Omega_r(C,t_f)} \;.
    \label{expC}
\end{equation}

An explicit expression for $\Omega_r(C,t_f)$ is obtained from a renewal equation. The idea is to consider the evolution from the start of the process until either one of two things happens. The first possibility is that the particle resets for the first time at some time $0\le t\le t_f$, after which the entire process restarts from the origin, with the remainder of the trajectory lasting a time $t-t_f$ and incurring a cost $C-c(x)$, where $c(x)\ge0$ is the cost of resetting from the point $x$ to the origin.  In this case, the particle resets with probability ${\rm e}^{-rt}r{\rm d}t$ in the interval $[t,t+{\rm d} t]$, and is distributed over space as $G(x;t)$ which is the Green function for diffusion,
\begin{equation}
    G(x;t) = \frac{1}{\sqrt{4\pi D t}}e^{-\frac{(x-x_0)^2}{4 D t}} \label{G0}
\end{equation}
where $D$ is the diffusion constant. The second possibility is the particle reaches the point $x$ at time $t_f$ without resetting. This event arises with probability ${\rm e}^{-rt}$, incurs zero cost and we allow only those endpoints $x$ that fall within the set $E$. For the case of RBM, $E$ is the entire real line, whereas for RBB, $E$ is the origin.

Expressing these two possibilities as a renewal equation, we find
\begin{widetext}
\begin{equation}
    \Omega_r(C,t_f) = \int_0^{t_f} {\rm d} t\, r{\rm e}^{-r t} \int_{-\infty}^\infty {\rm d} x\, G(x;t)
    \Omega_r(C-c(x),t_f-t) + {\rm e}^{-r t} \delta(C) \int_E {\rm d}x \, G(x;t_f) \;.\label{renew1} 
\end{equation}
\end{widetext}
The first term is a convolution and thus the recursion can be solved by introducing the double Laplace transform
\begin{equation}
\widetilde{\Omega}_r(p,s) = \int_0^\infty {\rm d} C\, {\rm e}^{-pC} \int_0^\infty {\rm d} t_f\, e^{-st_f} \Omega_r(C,t_f)\;.
\end{equation}

\paragraph*{Note on notation} Here we have used $\widetilde{\Omega}_r(N,p|x_0,s)$ to indicate a double Laplace transform. We will also use a tilde symbol to denote single Laplace transforms of a single time variable, e.g. $\widetilde{G}(x;s)$, below. The arguments of the function should  make  clear the number of Laplace variables. In certain places, for convenience, we will use ${\mathcal L}_{t\to s}$ to indicate Laplace transform to Laplace variable $s$ and ${\mathcal L}^{-1}_{s\to t}$ to indicate Laplace inversion to the time domain.
\medskip

Laplace transforming (\ref{renew1}) with respect to both arguments and rearranging, we find
\begin{equation}
    \widetilde{\Omega}_r(p,s) = \frac{\widetilde{K}(r+s)}{1 - r \widetilde{W}(p,r+s)}
\end{equation}
where
\begin{align}
    \label{K}
    \widetilde{K}(s) &=  \int_E {\rm d}x \, \int_{0}^{\infty}{\rm d}t_f\,{\rm e}^{-s t_f} G(x;t_f) =  \int_E {\rm d}x \, \widetilde{G}(x;s) \\
    \widetilde{W}(p,s) &= \int_{-\infty}^\infty {\rm d} x\, {\rm e}^{-p c(x)} \int_{0}^{\infty}{\rm d}t\, {\rm e}^{- s t} G(x;t)\nonumber \\  &= \int_{-\infty}^\infty {\rm d} x\, {\rm e}^{-p c(x)} \widetilde{G}(x;s)
\end{align}
and we have from (\ref{G0}) that
\begin{equation}
        \widetilde{G}(x,s) = \frac{1}{2\sqrt{Ds}}e^{-\sqrt{\frac{s}{D}}\abs{x}} \;.
\end{equation}
We see that the function $\widetilde{K}(s)$ is determined by the constraint placed on the endpoint of the trajectory, and that $\widetilde{W}(p,s)$ depends on the functional form of the cost. Once these functions have been determined for the cases of interest, we can obtain the mean cost as a function of $t_f$ from (\ref{expC}) via
\begin{equation}
    \expval{C} = \frac{ {\mathcal L}^{-1}_{s\to t_f} \{ - \partial_p \widetilde{\Omega}_r(p,s) |_{p\to 0^+} \}}{ {\mathcal L}^{-1}_{s\to t_f} \{ \widetilde{\Omega}_r(p,s)|_{p\to 0^+} \} } \;.
    \label{Cinv}
\end{equation}
Taking the limit $p\to0^+$, we find that the functions to be inverted to obtain the numerator and denominator, respectively, are
\begin{align}
\label{numinv}
\left. - \partial_p \widetilde{\Omega}_r(p,s) \right|_{p\to 0^+} = &\frac{r(r+s)^2}{s^2} \widetilde{K}(r+s)\times \nonumber\\  &\int_{-\infty}^\infty {\rm d}x\, c(x) \widetilde{G}(x;r+s) \\
\left. \widetilde{\Omega}_r(p,s) \right|_{p\to 0^+} = &\frac{r+s}{s} \widetilde{K}(r+s) \;.
\end{align}

\subsection{Resetting Brownian Motion (RBM)}

For the case of RBM, the trajectory endpoint is unconstrained, and the integral in (\ref{K}) is over all $x$. For any properly normalized distribution of endpoints we then have $\widetilde{K}(s) = 1/s$ and the denominator in (\ref{Cinv}) is unity. It thus remains to compute the numerator by inverting (\ref{numinv}) for the cost function of interest. In the main text we consider power-law cost functions,
\begin{equation}
    c(x) = 2^{n/2} \abs{y}^n \quad\mbox{where}\quad y = \frac{x}{\sqrt{2Dt_f}} \;,
\label{costn}
\end{equation}
specifically the linear and quadratic cases, $n=1$ and $n=2$. Substituting into (\ref{numinv}) yields
\begin{align}
\label{Crbm}
    \expval{C}_n^{\rm RBM} &= \frac{\Gamma(n+1)}{t_f^{n/2}} {\mathcal L}^{-1}_{s\to t_f} \left\{ \frac{r}{s^2} \frac{1}{(r+s)^{n/2}} \right\} \nonumber \\ &= \frac{\Gamma(n+1)}{R^{n/2}} \frac{R \gamma(\frac{n}{2},R) - \gamma(\frac{n}{2}+1, R)}{\Gamma(\frac{n}{2})}
\end{align}
where $R=r t_f$ is the dimensionless resetting rate, and $\gamma(s,x)$ is the lower incomplete Gamma function,
\begin{equation}
    \gamma(s,x) = \int_0^x {\rm d} u\, u^{s-1} {\rm e}^{-u} \;.
\end{equation}
Note that this result is obtained by recognizing the Laplace transform in (\ref{Crbm}) as the convolution of $t$ with $t^{n/2-1} {\rm e}^{-r t}/\Gamma(n/2)$, evaluated at $t_f$. Note further that $\langle C \rangle_0^{\rm RBM} = R$ and that for the special cases $n=1$ (`lin') and $n=2$ (`quad') considered in the main text, we have
\begin{align}
    \label{Crbmlin}
    \expval{C}^{\rm RBM}_{\rm lin} &= \frac{{\rm e}^{-R}}{\sqrt{\pi}} + 
    \frac{(2R-1)\erf\left(\sqrt{R}\right)}{2 \sqrt{R}} \\
    \label{Crbmquad}
    \expval{C}^{\rm RBM}_{\rm quad} &= \frac{2 \left(R+{\rm e}^{-R}-1\right)}{R} \;,
\end{align}
where we have used
$\Gamma(1/2) = \sqrt{\pi}$ ,
$\gamma(1/2, x) = \sqrt{\pi} \erf(\sqrt{x})$ and 
$\gamma(1, x) = 1 - {\rm e}^{-x}$
along with the recursion relation
\begin{equation}
    \gamma(s+1, x) = s \gamma(s, x) - x^s {\rm e}^{-x} \;.
\end{equation}

\subsection{Resetting Brownian Bridge (RBB)}

For the RBB, the calculation is a little more complex due to the constraint on the trajectory endpoint. Taking $E$ to comprise just the point at the origin in (\ref{K}) we now have $\widetilde{K}(s) = 1/\left(2\sqrt{Ds}\right)$ and the denominator of (\ref{Cinv}) takes the form
\begin{align}
\label{norm_time}
{\mathcal L}^{-1}_{s\to t_f} \{ \widetilde{\Omega}_r(p,s)|_{p\to 0} \} &= \frac{1}{2\sqrt{D}} {\mathcal L}^{-1}_{s\to t_f} \left\{ \left(1+\frac{r}{s}\right) \frac{1}{\sqrt{r+s}} \right\} \nonumber \\ &= 
\frac{{\rm e}^{-r t_f} + \sqrt{\pi r t_f} \erf(\sqrt{r t_f})}{2\sqrt{\pi D t_f}} \;.
\end{align}
Again, the inversion can be performed by recognizing it as a convolution.

Turning now to the numerator, we find for $c(x)$ given by (\ref{costn}) that
\begin{align}
    {\mathcal L}^{-1}_{s\to t_f} \{ - &\partial_p \widetilde{\Omega}_r(p,s) |_{p\to 0} \} = \nonumber \\ &\frac{1}{2\sqrt{D}} \frac{\Gamma(n+1)}{t_f^{n/2}}\times {\mathcal L}^{-1}_{s\to t_f} \left\{ \frac{r}{s^2} \frac{1}{(r+s)^{(n-1)/2}} \right\} \;.
\end{align}
Comparing with (\ref{Crbm}) we see that
\begin{equation}
    {\mathcal L}^{-1}_{s\to t_f} \{ - \partial_p \widetilde{\Omega}_r(p,s) |_{p\to 0} \} = \frac{n}{2\sqrt{Dt_f}} \expval{C}_{n-1}^{\rm RBM}
\end{equation}
and by dividing by the denominator (\ref{norm_time}) we find
\begin{equation}
    \expval{C}_{n}^{\rm RBB} = \frac{n\sqrt{\pi} \expval{C}_{n-1}^{\rm RBM}}{{\rm e}^{-R} + \sqrt{\pi R} \erf(\sqrt{R})} \;,
\end{equation}
recalling that $R=rt_f$. For the special cases $n=1$ and $n=2$ we have
\begin{align}
    \label{Crbblin}
    \expval{C}^{\rm RBB}_{\rm lin} &= \frac{\sqrt{\pi}R}{{\rm e}^{-R}+\sqrt{\pi R} \erf(\sqrt{R})} \\
    \label{Crbbquad}
    \expval{C}^{\rm RBB}_{\rm quad} &=2 - \sqrt{\frac{\pi}{R}} \frac{\erf (\sqrt{R})}{{\rm e}^{-R}+\sqrt{\pi R}\erf(\sqrt{R})} \;.
\end{align}

\medskip

We compare the functional forms of the mean cost between the two ensembles in Fig.~\ref{fig:costs}. Fig.~\ref{fig:linear_costs} shows how the mean linear cost behaves with $R$, and is seen to increase indefinitely as $R \to \infty$ as  $\sqrt{R}$. Corresponding plots for the mean quadratic cost are in Fig. \ref{fig:quad_costs}. Unlike the case for the linear cost, the mean total quadratic cost saturates as $R \to \infty$.

\begin{figure}[tb]
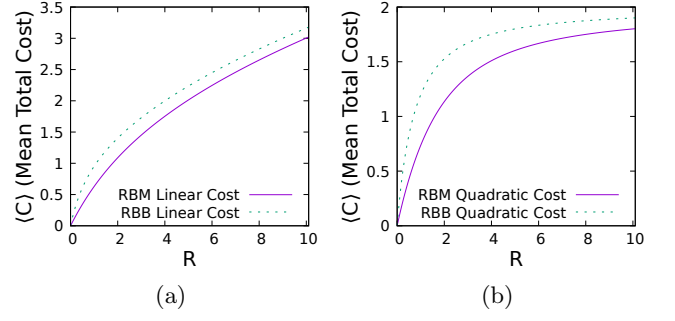


\centering
    \begin{subfigure}{.235\textwidth}
		\centering
	    \includegraphics[width=\linewidth]{Linear_cost_RBM_RBB.pdf} 
		\caption{}
        \label{fig:linear_costs}
	\end{subfigure}
    \begin{subfigure}{.235\textwidth}
		\centering
	    \includegraphics[width=\linewidth]{Quadratic_cost_RBM_RBB.pdf} 
		\caption{}
        \label{fig:quad_costs}
	\end{subfigure}
 \caption{Comparison of mean total cost $\langle C \rangle$ for (a) linear cost and (b) quadratic cost per reset for RBM and RBB. It can be seen that the costs increase monotonically for both the cases. This is consistent with the intuition that more frequent resets will incur a higher cost. Although in the case of a linear cost the mean cost increases indefinitely with $R$ we find that in the case of  a quadratic cost the mean cost saturates as $R\to \infty$.}
 \label{fig:costs}
\end{figure}

\section{Derivation of Success Probabilities}
\label{sec:sprob}

\subsection{Resetting Brownian Motion (RBM)}

For a Resetting Brownian Motion, the probability of reaching a target at position $m$ by time $t_f$ can be obtained from the target's survival probability which was given as Eq.~(6) in \cite{EM11a}. Noting that the target surviving corresponds to an unsuccessful search, we find that the success probability is given by the inverse Laplace transform
\begin{align}
    P_s^{\text{RBM}} = \int_{\Gamma} \frac{{\rm d}s}{2\pi i}{\rm e}^{st_f} \frac{1}{s} \frac{r+s}{r+s{\rm e}^{\sqrt{\frac{r+s}{D}}m}}\;.
\end{align}
where $\Gamma$ is the Bromwich contour. By introducing rescaled variables $u = st_f, R = rt_f$ and $M = m/\sqrt{2Dt_f}$, we obtain the form presented in the main text,
\begin{align}
    \label{Prbm}
    P_s^{\text{RBM}} = \int_{\Gamma} \frac{{\rm d}u}{2\pi i}e^{u} \frac{1}{u} \frac{R+u}{R+u {\rm e}^{M\sqrt{2}\sqrt{u+R}}}\;.
\end{align}

\subsection{Resetting Brownian Bridge (RBB)}

The success probability for a Resetting Brownian Bridge is provided as Eq.~(54) in the Supplemental Material of Ref.~\cite{BMS22}. In the notation of the present work, this reads
\begin{align}
    \label{Prbb}
    P_s^{\text{RBB}} = \phi(R) \int_\Gamma \frac{{\rm d} u}{2\pi i}{\rm e}^u \frac{\sqrt{u+R}}{u} \frac{R+u {\rm e}^{-M\sqrt{2}\sqrt{u+R}}}{R+u {\rm e}^{M\sqrt{2}\sqrt{u+R}}}\;,
\end{align}
where
\begin{align}
    \phi(R) = \frac{\sqrt{\pi}}{e^{-R}+\sqrt{\pi R}\erf\left( \sqrt{R} \right)}\;.
\end{align}
We can plot the success probabilities (\ref{Prbm}) and (\ref{Prbb}) by performing the inverse Laplace transforms numerically \cite{AW06}. In Figure~\ref{fig:ps_m_1}, we plot the two functions obtained at fixed $M=1$. For RBM, we find that the success probability decreases monotonically with $R$, whilst for RBB, the function is peaked at some nonzero resetting rate $R$. When $M$ is reduced to $1/2$, Figure~\ref{fig:ps_m_0_5}, we find that both functions are peaked.
\begin{figure}[tb]
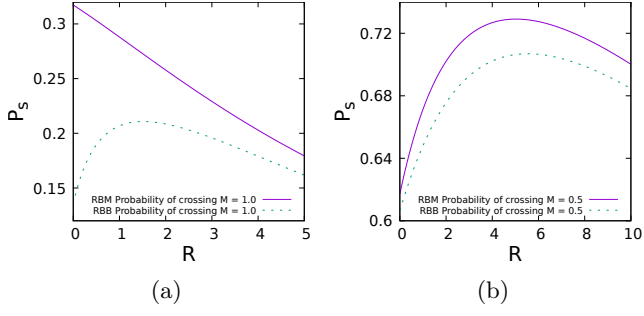

\centering
    \begin{subfigure}{.235\textwidth}
		\centering
	    \includegraphics[width=\linewidth]{Ps_m_1_RBM_RBB.pdf} 
		\caption{}
        \label{fig:ps_m_1}
	\end{subfigure}
    \begin{subfigure}{.235\textwidth}
		\centering
	    \includegraphics[width=\linewidth]{Ps_m_0_5_RBM_RBB.pdf} 
		\caption{}
        \label{fig:ps_m_0_5}
	\end{subfigure}
\caption{Comparison of probability of finding the target $(P_{\rm s})$ as a function $R$ for (a) $M = 1.0$ and (b) $M = 0.5$. For RBM (purple), $P_{\rm s}$ can initially be a decreasing of increasing function function depending on the value of $M$, whereas for RBB (green) $P_{\rm s}$ always increases initially with resetting, reaches a maximum then decreases for large $R$.}
 \label{fig:ps}
\end{figure}
\\

\section{Success Probability Expansions} \label{appendix_landau}

The expansion of $P_{\rm s}$ for  the two resetting ensembles can be obtained by expanding out the integrands of \eqref{phit_RBM} and \eqref{phit_RBB} in terms of $R$ and integrating term by term.To perform these expansions, it is helpful to define $s = u+R$, so that they become
\begin{align}
    P_{\text{s}}^{\text{RBM}} = &{\rm e}^{-R}\int_\Gamma \frac{{\rm d}u}{2\pi i}{\rm e}^u  \frac{1}{s-R} \frac{s}{R+(s-R){\rm e}^{M\sqrt{2s}}}\;, \label{p_hit_RBM2}\\
    P_{\text{s}}^{\text{RBB}} = &\frac{\sqrt{\pi}{\rm e}^{-R}}{\sqrt{\pi R}\erf \left( \sqrt{R} \right)+{\rm e}^{-R}} \times\nonumber\\ &\int_\Gamma \frac{ds}{2\pi i} {\rm e}^s  \frac{\sqrt{s}}{s-R} \frac{R+(s-R) {\rm e}^{-M\sqrt{2s}}}{R+(s-R) e^{M\sqrt{2s}}}\;. \label{p_hit_RBB2}
\end{align}
The procedure now is to expand these expressions as a power series in $R$, and invert term-by-term. Up to second order, we find for the RBM case the expansion
\begin{align}
     P^{\text{RBM}}_{\text{s}} = & g_0(M) + g_1(M)R + g_2(M)\frac{R^2}{2}+ \ldots \;, \label{phit_expansion_RBM_S}
\end{align}
where the coefficients $g_i(M)$ are given by the integrals
\begin{widetext}
\begin{align}
    g_0(M) = &\int_\Gamma \frac{{\rm d}s}{2\pi i} {\rm e}^s \frac{{\rm e}^{-\sqrt{2} M \sqrt{s}}}{s} = \text{erfc}\left(\frac{M}{\sqrt{2}}\right) \\[1ex]
    g_1(M) = &\int_\Gamma \frac{{\rm d}s}{2\pi i} {\rm e}^s \frac{{\rm e}^{-2 \sqrt{2} M \sqrt{s}} \left(-1-{\rm e}^{\sqrt{2} M \sqrt{s}} (-2+s)\right)}{s^2} \\[1ex]
   g_2(M) = &\int_\Gamma \frac{{\rm d}s}{2\pi i} {\rm e}^s \frac{{\rm e}^{-3 \sqrt{2} M \sqrt{s}} \left(2+2 {\rm e}^{\sqrt{2} M \sqrt{s}} (-3+s)+{\rm e}^{2 \sqrt{2} M \sqrt{s}} \left(6-4 s+s^2\right)\right)}{s^3} \;,
\end{align}

each of which can, like $g_0(M)$, be expressed in a closed form in terms of error functions. Explicit expressions are provided in an accompanying Mathematica notebook uploaded to DataShare \cite{link}.

Similarly for RBB, we have
\begin{align}
    P^{\text{RBB}}_{\text{s}} = & h_0(M) + h_1(M)R + h_2(M)\frac{R^2}{2} + \ldots \;,  \label{phit_expansion_RBB_S}
\end{align}
where

\begin{align}
    h_0(M) =& \int_\Gamma \frac{{\rm d}s}{2\pi i}{\rm e}^s \frac{{\rm e}^{-2 \sqrt{2} M \sqrt{s}} \sqrt{\pi }}{\sqrt{s}} = {\rm e}^{-2 M^2}\;,\\[1ex]
    h_1(M) = &\int_\Gamma \frac{{\rm d}s}{2\pi i} {\rm e}^s \frac{{\rm e}^{-3 \sqrt{2} M \sqrt{s}} \sqrt{\pi } \left(-1+{\rm e}^{2 \sqrt{2} M \sqrt{s}}+{\rm e}^{\sqrt{2} M \sqrt{s}} (1-2 s)\right)}{s^{3/2}} \;,\\[1ex]
   h_2(M) = &\int_\Gamma \frac{{\rm d}s}{2\pi i}e^s \frac{2 {\rm e}^{-4 \sqrt{2} M \sqrt{s}} \sqrt{\pi } \left(3+6 {\rm e}^{\sqrt{2} M \sqrt{s}} (-1+s)-6 {\rm e}^{3 \sqrt{2} M \sqrt{s}} (-1+s)+2 {\rm e}^{2 \sqrt{2} M \sqrt{s}} s (-3+4s)\right)}{3 s^{5/2}}
\end{align}

which again have closed-form expressions.

A similar process can be used to obtain the terms beyond the quadratic term. Note that in $R \to 0$ limit, \eqref{phit_expansion_RBM_S} and \eqref{phit_expansion_RBB_S} reduce to known results without resetting:  $P^{\text{RBM}}_{\text{s}} = \text{erfc}\left(M/\sqrt{2}\right)$ \cite{Redner} and $P^{\text{RBB}}_{\text{s}} = {\rm e}^{-2M^2}$ \cite{BMS22}.

The coefficients $a_i$ in \eqref{landau_expansion} can be written in terms of the terms of expansion of $P_{\text{S}}$ as
\begin{align}
    a_0 = &C_0 -\lambda  \log \left(g_0\right)\;,\\
    a_1 = &C_1-\lambda\frac{g_1}{g_0}\;,\\
    a_2 = &\frac{C_2}{2}-\lambda\frac{\left(-g_1^2+g_0 g_2\right)}{2g_0^2}\;,\\
    a_3 = &\frac{C_3}{6}-\lambda\frac{  \left(2 g_1^3-3 g_0 g_1 g_2+g_0^2 g_3\right)}{6 g_0^3}\;,\\
    a_4 = &\frac{C_4}{24}-\lambda\frac{  \left(-6 g_1^4+12 g_0 g_1^2 g_2-3 g_0^2 g_2^2-4 g_0^2 g_1 g_3+g_0^3 g_4\right)}{24 g_0^4}\;,
\end{align}
where $C_n = d^n \expval{C}/dR^n\big\vert_{R\to0^+}$. The $g_i$ should be substituted with $h_i$ in the RBB case.
\end{widetext}
\bibliography{main}

\end{document}